\documentclass[%
 reprint,
 amsmath,amssymb,
 aps,
]{revtex4-1}

\usepackage{dcolumn}
\usepackage{bm}

\usepackage{amsmath, amssymb}
\usepackage{amsfonts}
\usepackage{braket}

\usepackage[dvips]{graphicx}

\usepackage{ascmac}

\usepackage{comment}

\usepackage{color}

\usepackage{multirow}

\newcommand{\todayd}{\the\year/\the\month/\the\day}
\newcommand{\eq}[1]{\begin{equation} #1 \end{equation}}
\newcommand{\eqa}[2]{\begin{equation} #1 \label{#2} \end{equation}}
\newcommand{\del}{\partial}
\newcommand{\ep}{\epsilon}
\newcommand{\la}{\langle}
\newcommand{\ra}{\rangle}

\newcommand{\bib}{\bibitem}

\newcommand{\lr}{\leftrightarrow}
\newcommand{\const}{\mathrm{const}}

\newcommand{\balign}[1]{\begin{align} #1 \end{align}}

\newcommand{\lb}{\label}
\newcommand{\nt}{\notag}

\newcommand{\eref}[1]{Eq.~\eqref{#1}}

\newcommand{\fref}[1]{Fig.~\ref{f:#1}}

\newcommand{\ft}[2]{\left. #1 \right|_{#2}}

\DeclareMathOperator*{\argmax}{arg\,max}

\newcommand{\erase}[1]
{
#1
}

\newcommand{\figin}[4]
{
\erase{
\begin{figure}[tb]\centering\includegraphics[width= #1]{#2}\caption{#3}\label{f:#4}\end{figure}
}
}

\newcommand{\figinb}[4]
{
\erase{
\begin{figure}[b]\centering\includegraphics[width= #1]{#2}\caption{#3}\label{f:#4}\end{figure}
}
}

\def \({\left(}
\def \){\right)}

\def\rnum#1{\resizebox{0.5em}{\height}{\expandafter{\romannumeral #1}}}
\def\Rnum#1{\resizebox{0.5em}{\height}{\uppercase\expandafter{\romannumeral #1}}}

\begin{document}

\preprint{APS/123-QED}

\title{Attainability of Carnot Efficiency with Autonomous Engines}

\author{Naoto Shiraishi}
\affiliation{%
Department of Basic Science, The University of Tokyo, \\
3-8-1 Komaba, Meguro-ku, Tokyo 153-8902, Japan
}%
\date{\today}

\begin{abstract}
The maximum efficiency of autonomous engines with finite chemical potential difference is investigated.
We show that without a particular type of singularity autonomous engines cannot attain the Carnot efficiency.
This singularity is realized in two ways: single particle transports and the thermodynamic limit.
We demonstrate that both of two ways really lead to the Carnot efficiency in concrete setups.
Our results clearly illustrate that the singularity plays a crucial role for the maximum efficiency of autonomous engines.

\begin{description}
\item[PACS numbers]
05.70.Ln, 05.40.-a, 87.10.Mn, 87.16.Nn.
\end{description}
\end{abstract}

\pacs{Valid PACS appear here}
\maketitle

{\it Introduction.}
---The maximum efficiency of heat engines has been one of the central issues in thermodynamics.
Carnot showed that the efficiency of an engine attached to heat baths with temperature $T_H$ and $T_L$ ($T_H>T_L$) is bounded by $1-T_L/T_H$~\cite{Carnot}.
The upper bound is attained when the external control on the engine is quasistatic.
For the case with two particle baths under isothermal condition, the efficiency is bounded by $1$.
These maximum efficiencies are called the Carnot efficiency (CE).
Nonequilibrium thermodynamics has recently been applied to small fluctuating systems with external control, where the maximum efficiency analogous to macroscopic thermodynamics has been established~\cite{Sekimotobook}.

Since most engines from electric power plants to molecular motors are autonomous, thermodynamics for autonomous engines is an important issue.
Here, the word {\it autonomous} stands for systems with no time-dependent control parameter in non-equilibrium steady states~\cite{autonomous}.
Since all variables of the engine inevitably fluctuate, the attainability of the CE with autonomous engines is a non-trivial problem.
In the linear response regime (i.e., $T_H-T_L\simeq 0$), it is well known that the tight-coupling condition is necessary for autonomous engines to attain the CE~\cite{Broeck3}.
On the other hand, for the case of finite difference of temperatures or chemical potentials, most of studies have paid attention to specific and elaborated models~\cite{Buttiker, Landauer, Feynman, Matsuo, Parrondo, Sekimoto97, Jarzynski, Hondou, Sekimoto, Esposito, SS, SIKS, Mahan, Humphrey, Horvat, Sokolov, Derenyi, Sanchez, Jordan, Seifert, MJ, autoMJ, SMS}.
Famous models of autonomous engines are Feynman's ratchet~\cite{Feynman} and the B\"{u}ttiker-Landauer system~\cite{Buttiker, Landauer}, which convert heat flux into work.
Although they seemingly attain the CE~\cite{Feynman, Matsuo}, it has been established that these models actually cannot attain the CE in physically plausible setups~\cite{Parrondo, Sekimoto97, Jarzynski, Hondou}.
Another famous model is an information engine~\cite{Sekimoto, Esposito, SS, SIKS}, which performs autonomous control and always transports a single particle between particle baths.
In contrast to Feynman's ratchet and the B\"{u}ttiker-Landauer system, the information engine attains the CE.
In addition, limiting effective filters of energy or chemical potential~\cite{Mahan, Humphrey, Horvat, Sokolov, Derenyi}, a quantum dot~\cite{Sanchez, Jordan}, and soft nanomachines~\cite{Seifert} also attain the CE.
However, contrary to externally-controlled engines, the comprehensive understanding of autonomous engines with finite difference of temperatures or chemical potentials has been elusive.
Especially, the understanding of macroscopic autonomous engines is missing.

In this Rapid Communication, we address the issue of the general condition for autonomous engines to attain the CE with finite chemical potential difference. 
To demonstrate this, we first introduce a schematic model; an autonomous version of the macroscopic Carnot engine.
This model clearly illustrates the characteristics of autonomous engines that in normal nonsingular setups they cannot attain the CE even with infinitely slow dynamics.
Contrary to this, in the case of singular transition rates, since this singularity prohibits the particle leakage from the dense bath to the dilute bath, this engine attains the CE.
We then move to general discussion and prove that without a special type of singularity any autonomous engine cannot attain the CE, which is consistent with both our model and the existing models~\cite{Buttiker, Landauer, Feynman, Matsuo, Parrondo, Sekimoto97, Jarzynski, Hondou, Sekimoto, Esposito, SS, SIKS, Mahan, Humphrey, Horvat, Sokolov, Derenyi, Sanchez, Jordan, Seifert, MJ, autoMJ, SMS}.

\figin{8.5cm}{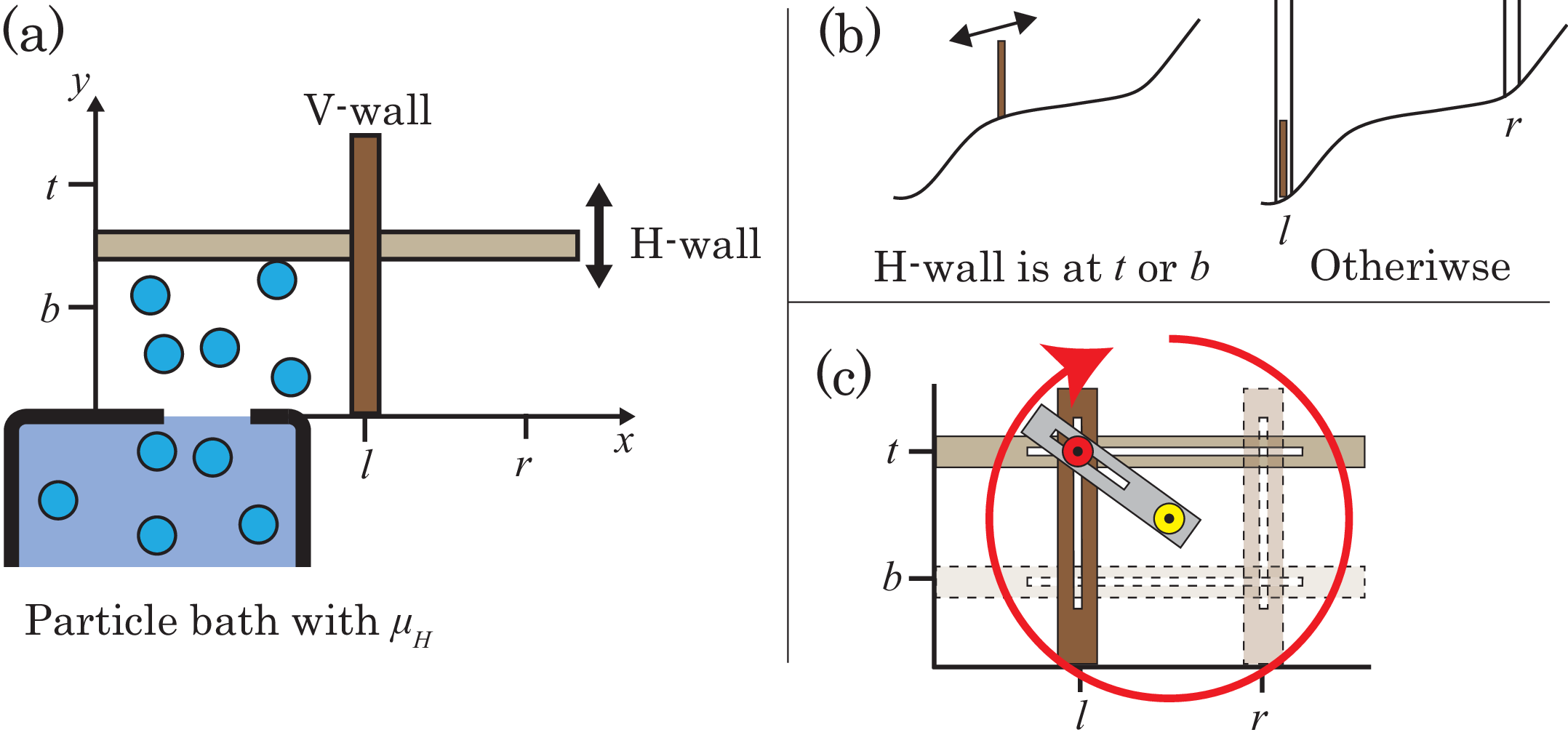}
{(Color online):
(a) Schematic of the autonomous Carnot engine, which consists of V-wall and H-wall.
Since V-wall is at $l$, particles can be exchanged with the particle bath with $\mu _H$.
(b) The potential landscape of V-wall.
If H-wall is at $t$ or $b$, V-wall is movable.
Otherwise, V-wall is trapped at $l$ or $r$ by delta-function type potentials.
(c) Schematic of the torque.
The red disk is on the cross point of V-wall and H-wall, which slides on both walls.
The yellow disk is fixed and serves as a shaft of the gray slat.
With a single rotation $A\to B\to C\to D\to A$, the gray slat makes one rotation and we extract mechanical work.
}{system}

{\it Kinetic model and its coarse-graining.}
---It is hard for autonomous engines to attain the CE, and they attain the CE only if they possess a special type of singularity.
To demonstrate the above characteristics of autonomous engines, we introduce an autonomous Carnot engine, which extracts mechanical work from particle flux between particle baths with given chemical potentials $\mu _H$ and $\mu _L$.
The engine consists of two movable walls; V-wall and H-wall (see \fref{system}.(a)).
Only when H-wall (V-wall) is at the position $t$ or $b$ ($l$ or $r$), V-wall (H-wall) can move along the $x$ ($y$) axis, and otherwise V-wall (H-wall) is fixed at $l$ or $r$ ($t$ or $b$) (see \fref{system}.(b)).
Thus, the engine has four stable positions, $(l,b)$, $(l,t)$, $(r,t)$, and $(r,b)$, which we denote by $A$, $B$, $C$, and $D$, respectively.
When V-wall is at the position $l$ ($r$), the engine can exchange particles with the bath with $\mu _H$ ($\mu _L$).
Otherwise, the engine cannot exchange particles.
The engine is under isothermal condition, and the dynamics of the walls and particles are stochastic (the explicit time-evolution equation is shown in the Supplemental Material).
Since the cross point of two walls passes through the rectangular-shaped trajectory $A\to B\to C\to D\to A$, by imposing torque on the cross point we extract mechanical work automatically (see \fref{system}.(c) and \fref{cycle}).

We here adopt a well-used approximation that with given position of the walls and particle number the particles are in equilibrium~\cite{Cerino, Sano}.
If the equilibration of particles is much faster than the dynamics of the walls and the exchange of particles with the baths, the above approximation is justified.
This time separation leads to the coarse-grained description; the Markov jump processes with discrete states $(X,n)$, where $X\in \{ A,B,C,D\}$ represents the position of the walls and $n$ represents the particle number.
We call this model as coarse-grained autonomous Carnot engine (CGACE).

The energy difference from $A$ to $B$ including the external force is denoted by $E_{AB}$, and $E_{BC}$, $E_{CD}$, $E_{DA}$ are defined in a similar manner.
Then the work against the imposed force per one rotation $A\to B\to C\to D\to A$ is written as $E_{AB}+E_{BC}+E_{CD}+E_{DA}=:W_{\rm tot}>0$.
Since the engine is under isothermal condition, the transition rates of V-wall (i.e., $A\lr D$ and $B\lr C$) should satisfy the local detailed balance condition:
\eqa{
\ln \frac{P(X\to X^-;n)}{P(X^-\to X;n)}=-\beta (E_{XX^-}+F(V_{X^-},n)-F(V_X,n)),
}{ldb}
where we defined $A^-:=D, B^-:=C, C^-:=B, D^-:=A$, and $E_{XX^-}:=-E_{X^-X}$, and $\beta =1/k_{\rm B}T$ is the inverse of the Boltzmann constant times temperature.
$P(X\to X^-;n)$, $V_X$, and $F(V,n)$ represent the transition rate from $(X,n)$ to $(X^-,n)$, the volume at $X$, and the Helmholtz free energy with volume $V$ and particle number $n$, respectively.
Note that the details of the transition rates of the H-wall as $P((A,n)\to (B,n'))$ are not important in the following discussion.

\figin{9cm}{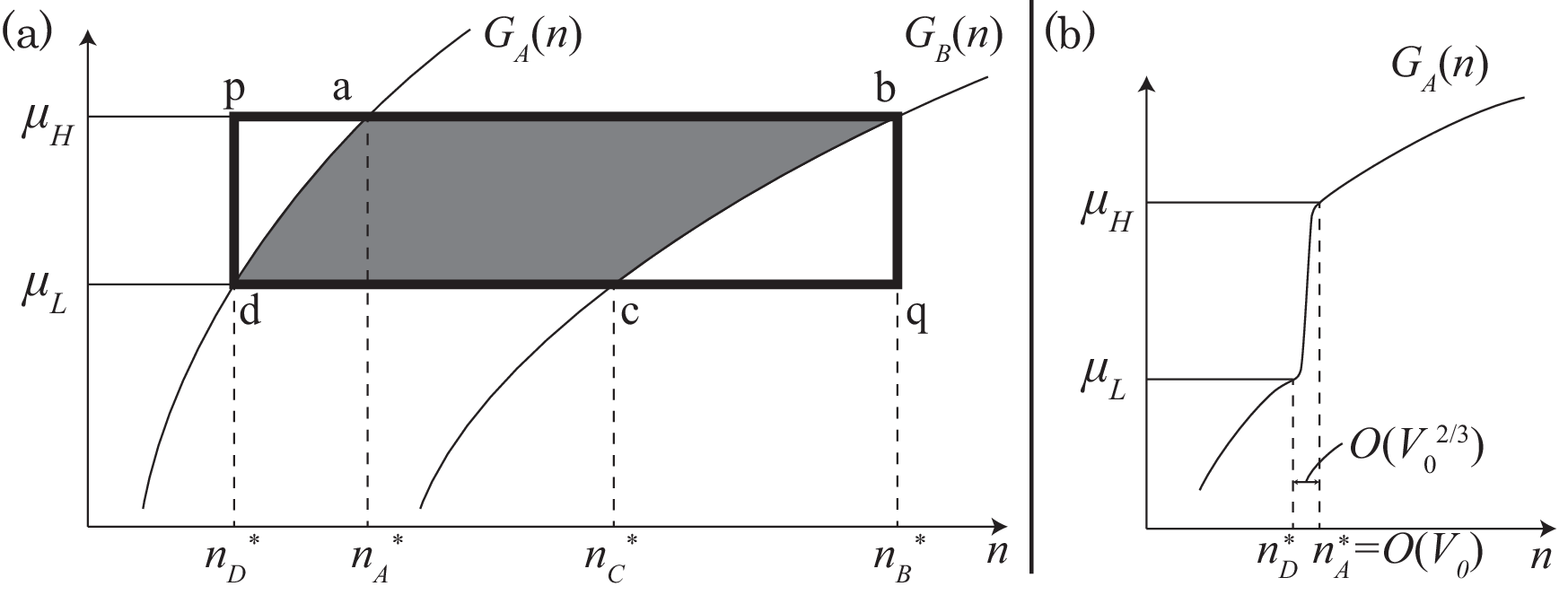}
{(a) Graphs of $G_A(n)$ and $G_B(n)$.
The area of ``abcd" (colored by gray) corresponds to the rhs of \eref{key}, which is the upper bound for $W_{\rm tot}$.
The area of ``pbqd" (surrounded by bold lines) corresponds to the lower bound for $C_\mu$.
(b) A graph of $G_A(n)$ with singular transition rates \eqref{pad} and \eqref{pda}.
$G_A$ shows almost discontinuous behavior, which allows $n_A^*$ and $n_D^*$ as $n_A^*-n_D^*=O({V_0}^{2/3})$.
}{GX}

{\it Maximum efficiency of CGACE.}
---To confirm the difficulty for autonomous engines to attain the CE, we here derive the maximum efficiency of the CGACE with fixed $\mu _H$ and $\mu _L$.
In the following, we investigate the condition for the maximum efficiency, and then calculate the efficiency under this condition.
First, the engine with the maximum efficiency should prevent two kinds of leakage.
One is leakage of particles: 
If the dynamics of H-wall is much slow, V-wall moves between $B\lr C$ or $A\lr D$ many times and particles leak from the dense bath to the dilute bath without extracting work.
The other is the leakage of energy: 
If the exchange of particles between the baths and the engine is much slow, the walls rotate obeying the external force as $A\to D\to C\to B\to A$ and the work is lost.
Hence, it is plausible that the maximum efficiency is realized when the dynamics of V-wall is much slow than that of H-wall and particles, and we treat this situation in the following.

The stationary distribution $P_{\rm st}(B, n)$, for example, is then calculated as
\eq{
P_{\rm st}(B,n)=P_{\rm st}(A,B)\frac{e^{-\beta (F(V_B,n)+E_{AB}-\mu _Hn)}}{Z_{AB}},
}
where $P_{\rm st}(A,B)$ represents the stationary probability at $A$ or $B$, and $Z_{AB}:=\sum_ne^{-\beta (F(V_A,n)-\mu _Hn)}+e^{-\beta (F(V_B,n)+E_{AB}-\mu _Hn)}$ is a normalization constant.
We denote the stationary probability flux of $X\to X^-$ with $n$ particles by $j_{X\to X^-}(n):=P_{\rm st}(X,n)P(X\to X^-;n)$.
Owing to the law of large numbers, the realized particle number when the transition $X\to X^-$ occurs is around $n_X^*:=\argmax _n j_{X\to X^-}(n)$.

We now calculate the efficiency $\eta :=W_{\rm tot}/C_\mu$, where $C_\mu$ represents the average consumption of chemical potential per a single rotation $A\to B\to C\to D\to A$.
The condition that the direction of the dominant dynamics of the walls is $A\to B\to C\to D\to A$ leads to
\balign{
j_{A\to D}(n_A^*)<&j_{D\to A}(n_D^*), \lb{jcond1} \\
j_{C\to B}(n_C^*)<&j_{B\to C}(n_B^*). \lb{jcond2}
}
Summing the logarithms of \eqref{jcond1} and \eqref{jcond2}, and using the local detailed balance condition \eqref{ldb}, we arrive at a key inequality:
\balign{
W_{\rm tot}
<& \( \mu _Hn_B^*-\mu _Ln_C^*-\int _{n_C^*}^{n_B^*}G_B(n)dn\) \nt \\
&-\( \mu _Hn_A^*-\mu _Ln_D^*-\int _{n_D^*}^{n_A^*}G_A(n)dn\) . \lb{key}
}
Here, $G_X(n)$ ($X=A, B$) is defined as
\balign{
G_X(n)&:=\frac{\del}{\del n}\( F(V_X,n)-\frac{1}{\beta}\ln P(X\to X^-;n)\) , 
}
and $n_X^*$ satisfies $G_A(n_A^*)=G_B(n_B^*)=\mu _H$ and $G_B(n_C^*)=G_A(n_D^*)=\mu _L$.
Note that the second law of thermodynamics implies monotonic increase of $G_X(n)$.
The right-hand side (rhs) of \eqref{key} corresponds to the area of ``abcd" (colored by gray) in \fref{GX}.(a).
In addition,  $C_\mu$ is evaluated as $C_\mu \geq (\mu _H-\mu _L)(n_B^*-n_D^*)$, whose rhs corresponds to the area of ``pbqd" (surrounded by bold lines) in \fref{GX}.(a).
If a finite constant $a<+\infty$ satisfies
\eq{
n\frac{\del G_A}{\del n}\leq a, \ \ n\frac{\del G_B}{\del n}\leq a \lb{acond}
}
for any $n$, the inequality \eqref{key} implies
\eq{
\eta :=\frac{W_{\rm tot}}{C_\mu} \leq \frac{a\( 1-e^{-\beta \Delta \mu /a}\)}{\beta \Delta \mu}<1=\eta _{\rm Carnot}, \lb{bound}
}
where we used $G_A(n)\leq a\ln (n/n_D^*)+\mu _L$ for $n\geq n_D^*$ and $G_B(n)\geq a\ln (n/n_B^*)+\mu _H$ for $n\leq n_B^*$.
The inequality \eqref{bound} indicates that the maximum efficiency of the CGACE is strictly less than the CE.
Especially, if the transition rates of V-wall obey the symmetric rule~\cite{sym} or the Arrhenius rule~\cite{Arrhe}, the condition \eqref{acond} is equivalent to the condition for the thermodynamic function of the gas: $\sup_{n,V}n\cdot {\del ^2F}/{\del n^2}\leq a$.
We note that the ideal gas satisfies $n\cdot\del ^2F/\del n^2=\beta$ for any $n$ and $V$.

In the foregoing discussion, it was shown that the CGACE cannot attain the CE with normal transition rates.
However, the CGACE attains the CE with the transition rate with a special type of singularity.
We again assume that the dynamics of V-wall is much slow comparing to the H-wall and the particle exchange.
We set the transition rates between $A$ and $D$ as
\balign{
P(A\to D;n)&=\frac{k\cdot e^{-\beta (F(V_D,n)-E_{DA})}}{e^{-\beta F(V_A,n)}+e^{-\beta (F(V_D,n)-E_{DA})}}, \lb{pad} \\
P(D\to A;n)&=\frac{k\cdot e^{-\beta F(V_A,n)}}{e^{-\beta F(V_A,n)}+e^{-\beta (F(V_D,n)-E_{DA})}}, \lb{pda}
}
with a constant $k$.
The transition rates $P(B\to C;n)$ and $P(C\to B;n)$ are also written in a similar manner with the same $k$.
We note that such transition rates are physically realizable (see the Supplemental Material).

The crucial point of the form of $P(A\to D;n)$ is that $G_A$ shows an almost discontinuous jump from $\mu _L$ to $\mu _H$ (see \fref{GX}.(b)).
This discontinuity leads to the divergence of $n\del G_A/\del n$, and thus the left side of the inequality \eref{bound} does not prohibit the attainability of the CE.
We then properly set $V_A,V_B,V_C,V_D=O(V_0)$ and $E_{AB},E_{BC},E_{CD},E_{DA}=O(V_0)$ as satisfying $n_A^*-n_D^*=O({V_0}^{2/3})$ and $n_B^*-n_C^*=O({V_0}^{2/3})$, which are negligible in thermodynamic limit.
Under this setup, the efficiency is evaluated as
\eq{
\eta :=\frac{W_{\rm tot}}{C_\mu }\geq 1-O\( \frac{1}{{V_0}^{1/3}}\) ,
}
which indicates the attainability of the CE with thermodynamic limit $V_0\to \infty$  (detailed setups and calculations are discussed in the Supplemental Material).

\figin{8.8cm}{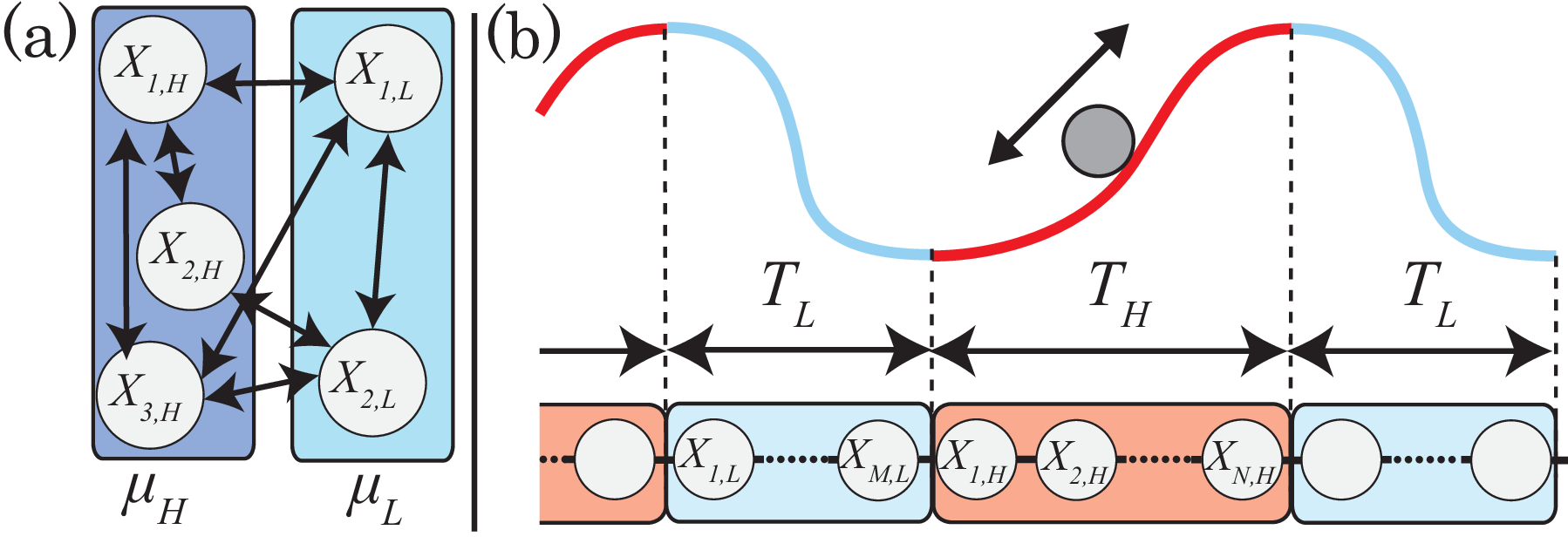}
{(Color online):
(a) An instance of the state space of an engine.
$X_{1,H}, X_{2,H}, X_{3,H}$ touches the particle bath with $\mu _H$, and $X_{1,L}, X_{2,L}$ touches the particle bath with $\mu _L$.
Arrows represent possible transitions.
(b) Schematic of the (discretized) B\"{u}ttiker-Landauer system and its state space.
$X_{i,H}$ ($X_{i,L}$) represents the position of the Brownian particle with a hot (cold) bath.
The whole state of the system is determined by the position $X$, the energy $E$, and the direction of the motion $c\in \{ +,-\}$.
}{gen}

{\it Necessary condition to attain the CE.}
---We now leave the specific model and go to the argument on general autonomous engines with finite chemical potential difference, which include models in Refs.~\cite{Buttiker, Landauer, Feynman, Matsuo, Parrondo, Sekimoto97, Jarzynski, Hondou, Sekimoto, Esposito, SS, SIKS,Mahan, Humphrey, Horvat,  Sokolov, Derenyi, Sanchez, Jordan, Seifert, MJ, autoMJ, SMS}.
We note that if the engine has continuous variables, we take their proper discretization.
We also note that our argument for particle baths is easily extended to the case of thermal baths.
By adding a new stochastic parameter if necessary (see the Supplemental Material), an autonomous engine satisfies the following two conditions:
(a) The engine touches at most one particle bath at one moment.
(b) The energy difference of the engine between two states (as $E_{XX^-}$ in the CGACE) is independent of its particle number $n$.
We then divide the possible states of the engine into $\{ X_{1,H},\cdots ,X_{n,H}\}$ and $\{ X_{1,L},\cdots ,X_{m,L}\}$, where a state $X_{i,H}$($X_{i,L}$) is attached to the bath with $\mu _H$($\mu _L$) (see \fref{gen}.(a) or \fref{cycle}).
The state of the whole system is written as $(X,n)$.
In the case of the CGACE, the engine takes four possible states $\{ A,B\} =\{ X_{1,H},X_{2,H}\}$ and $\{ C,D\} =\{ X_{1,L}, X_{2,L}\}$.
In the case of the (discretized) B\"{u}ttiker-Landauer system~\cite{Buttiker, Landauer, Matsuo, Hondou}, $X_H$($X_L$) corresponds to the position of the Brownian particle attached to a hot (cold) bath (see \fref{gen}.(b)).

We here use the fact that a thermodynamic engine which attains the CE satisfies the detailed balance condition~\cite{vanKampen}, which is a consequence of the widely believed conjecture that thermodynamic engines with the CE move quasistatically and has recently been proved for Markovian systems~\cite{mine}.
Then the stationary distribution for $n$ with given $X$ reduces to the grand canonical distribution.
Hence, the average of particle number transported from $X_{i,H}$ to $X_{j,L}$ per unit time is calculated as
\eqa{
\la n\ra _{X_{i,H}\to X_{j,L}} = \sum _n\frac{n\cdot f(n)e^{\beta\mu _Hn}}{\sum_{n'}f(n') e^{\beta\mu _Hn'}},
}{HtoL}
where we defined $f(n):=e^{-\beta F(X_{i,H},n)}P(X_{i,H}\to X_{j,L};n)$.
By assuming the local detailed balance condition for $X$, $\la n\ra _{X_{j,L}\to X_{i,H}}$ is calculated in a similar manner:
\eqa{
\la n\ra _{X_{j,L}\to X_{i,H}} = \sum _n\frac{n\cdot f(n)e^{\beta\mu _Ln}}{\sum_{n'}f(n') e^{\beta\mu _Ln'}}.
}{LtoH}
Here, we used the fact that the local detailed balance condition is written in a similar manner to \eref{ldb}, which follows from the condition (b).

Let $V_0$ be a typical system size.
Since the amount of extracted work is of order $V_0$, the necessary condition for absence of particle leakage between $X_{i,H}$ and $X_{j,L}$ is
\eq{
\frac{1}{V_0}\( \la n\ra _{X_{i,H}\to X_{j,L}}-\la n\ra _{X_{j,L}\to X_{i,H}} \) =0.
}
However, by compareing Eqs.~\eqref{HtoL} and \eqref{LtoH}, monotonic increase of $e^{\beta (\mu _H-\mu _L)n}$ in terms of $n$ yields 
\eq{
\frac{1}{V_0}(\la n\ra _{X_{i,H}\to X_{j,L}}-\la n\ra _{X_{j,L}\to X_{i,H}})\geq 0,
}
and the equality holds only when $f(V_0\rho )e^{\beta \mu _LV_0\rho}$ has a delta-function type singularity in terms of $\rho :=n/V_0$ such that:
\balign{
j _{X_{i,H}\to X_{j,L}}(V_0\rho )\propto f(V_0\rho )e^{\beta \mu _HV_0\rho}&\propto \delta (\rho -\rho ^*), \lb{sing-1} \\
j _{X_{j,L}\to X_{i,H}}(V_0\rho )\propto f(V_0\rho )e^{\beta \mu _LV_0\rho}&\propto \delta (\rho -\rho ^*). \lb{sing-2}
}
Here $j_{X\to X'}(n)$ represents probability flux of $X\to X'$ with particle number $n$.
Note that $\rho$ is not the particle density. 
The conditions \eqref{sing-1} and \eqref{sing-2} have a clear physical meaning: 
Particle numbers with both transitions $X_{i,H}\to X_{j,L}$ and $X_{j,L}\to X_{i,H}$ are always the same unique value $V_0\rho ^*$ within $o(V_0)$, which is the only way to prevent the leakage of particles.
Since $\mu_H$ and $\mu _L$ are fixed, this singularity appears in two ways: 
(i) $P(X_{i,H}\to X_{j,L};n)=0$ for all $n\in \mathbb{N}$ except $n=V_0\rho ^*$.
(ii) $1/V_0\cdot \del /\del \rho \ln f(V_0\rho )$ shows such discontinuity that
\balign{
\lim_{\rho ' \to \rho ^*-0}\lim_{V_0\to \infty}\ft{\frac{1}{V_0}\frac{\del}{\del \rho}\ln f(V_0\rho )}{\rho =\rho '} &\leq \mu _L, \\
\lim_{\rho ' \to \rho ^*+0}\lim_{V_0\to \infty}\ft{\frac{1}{V_0}\frac{\del}{\del \rho}\ln f(V_0\rho )}{\rho =\rho '} &\geq \mu _H.
}
To attain the CE, all possible transitions $X_{i,H}\lr X_{j,L}$ need to satisfy (i) or (ii).
Without such singularity, the engine cannot attain the CE as seen in \eref{bound}.
This is our main result.

Various existing autonomous engines which attains the CE~\cite{Sekimoto, Esposito, SS, SIKS, Matsuo, Seifert, Mahan, Humphrey, Horvat,  Sokolov, Derenyi, Sanchez, Jordan, MJ, autoMJ, SMS} adopt the way (i), where always a single particle or a particular value of energy is transported.
In this Rapid Communication, we construct an autonomous engine with the CE which adopts the way (ii).
In contrast, present autonomous engines which cannot attain the CE~\cite{Parrondo, Sekimoto97,Jarzynski, Hondou} satisfy neither (i) nor (ii).
In the case of the B\"{u}ttiker-Landauer system, for example, the momentum distributions of the hot bath and the cold bath at the contact point of these two baths are different.
Due to this difference, round-trips of the Brownian particle between two baths cause the energy transport from the hot bath to the cold bath in the form of kinetic energy, which implies finite dissipation.

{\it Concluding remarks.}
---In this Rapid Communication, we derived the necessary condition for autonomous engines to attain the CE.
The key property is a special type of singularity as Eqs.~\eqref{sing-1} and \eqref{sing-2}, which implies that always the same and unique amount of particle number or energy is transported between two states with different baths.
Without such singularities, an autonomous engine never attains the CE.
This result is consistent with existing results on the specific models of autonomous engines~\cite{Buttiker, Landauer, Feynman, Matsuo, Parrondo, Sekimoto97, Jarzynski, Hondou, Sekimoto, Esposito, SS, SIKS, Mahan, Humphrey, Horvat, Sokolov, Derenyi, Sanchez, Jordan, Seifert, MJ, autoMJ, SMS}.

Such singularities are realized only by the way of transports of a single particle or a particular energy (i) or a special type of thermodynamic limit (ii).
Previous models with the CE adopt the way (i), and no concrete model with the CE has adopted the way (ii).
In this Rapid Communication, we constructed a concrete example of the way (ii), and thus it was shown that both ways (i) and (ii) indeed lead to the CE.
Here, we should emphasize that the way (ii) is different from the following intuitive picture: 
For the case of the macroscopic heat engine, by decreasing the external force and slowing down the speed of the cyclic process, the engine will reach the CE.
In fact, this intuition is wrong for almost all macroscopic engines.
Although thermal fluctuation is usually negligible in macroscopic engines, in the case of infinitely slow speed thermal fluctuation should be taken into account even in macroscopic engines and the above intuition overlooks this inevitable fluctuation.
This fluctuation causes the leakage of particles or energy between two baths.
As evidence of this, the inequality \eqref{bound} still holds for the case of the infinitely slow speed.

Our results may give a new perspective on the physics of molecular motors.
Molecular motors are also autonomous engines, and in most cases they work with finite chemical potential difference.
Our results impose physical restrictions on molecular motors:
To attain the CE chemical heat engines should adopt the way of single particle transports (i) or thermodynamic limit (ii).
These two options look similar to the two types of molecular motors; working solely (like a kinesin) or collectively (like myosins in a muscle)~\cite{motorbook}.
It will be interesting if these two characteristics are connected.

{\it Acknowledgment.}
---The author thanks Takahiro Sagawa for fruitful discussions.
The author also thanks Eiki Iyoda, Yuki Izumida, Kyogo Kawaguchi, Tomohiko G Sano, and Akira Shimizu for helpful comments.
This work is supported by Grant-in-Aid for JSPS Fellows Number 26-7602.

\clearpage
\setcounter{equation}{0}
\def\theequation{S.\arabic{equation}}

\setcounter{figure}{0}
\def\thefigure{S.\arabic{figure}}

\begin{widetext}

\begin{center}
{\bf \Large Supplemental Material}

\

{\bf \large Time-evolution equation and dynamics of the walls}

\end{center}

We here write down the time-evolution equation of the walls explicitly.
Since V-wall and H-wall have the same form of time-evolution equations, we treat only the dynamics of V-wall.
The $x$-coordinate of the positions of V-wall, $l$, $r$, and $i$-th particle in the engine are denoted by $X$, $x_l$, $x_r$, and $x_i$, respectively.
The $y$-coordinate of the positions of H-wall, $t$, $b$ are denoted by $Y$, $y_t$, $y_b$, respectively.
Then the stochastic dynamics of V-wall is written as the overdamped Langevin equation:
\eq{
\gamma \frac{dX}{dt}=F_{{\rm ext},V}(X,Y)-\frac{dU_V(X)}{dX}-\frac{\del U_1(X,Y)}{\del X}-\sum _i\frac{\del U_2(x_i,X)}{\del X}+\sqrt{\frac{2\gamma}{\beta}}\xi _X.
}
Here, $\gamma$ represents the viscous coefficient, $F_{{\rm ext},V}(X,Y)$ represents the imposed external force on V-wall at $(X,Y)$, $U_V(X)$ represents the potential energy of V-wall, $U_1(X,Y)$ and $U_2(x_i,X)$ represent the interaction energy between V-wall and H-wall, and between V-wall and $i$-th particle, and $\xi _X$ represents the white Gaussian noise.
$U_1(X,Y)$ is given by
\balign{
U_1(X,Y):=&(\delta (X-x_l-\ep )+\delta (X-x_l+\ep )+\delta (X-x_r-\ep )+\delta (X-x_r+\ep )) \nt \\
&\cdot (\chi (y_t-\ep<Y<y_t+\ep)+\chi (y_b-\ep <Y<y_b+\ep )) \nt \\
&+(\delta (Y-y_t-\ep)+\delta (Y-y_t+\ep)+\delta (Y-y_b-\ep)+\delta (Y-y_b+\ep)) \nt \\
&\ \ \cdot (\chi (x_l-\ep<X<x_l+\ep)+\chi (x_r-\ep<X<x_r+\ep))
}
with sufficiently small $\ep$, where $\chi (P)$ takes one if the condition $P$ is true, and takes zero otherwise.

\figinb{17cm}{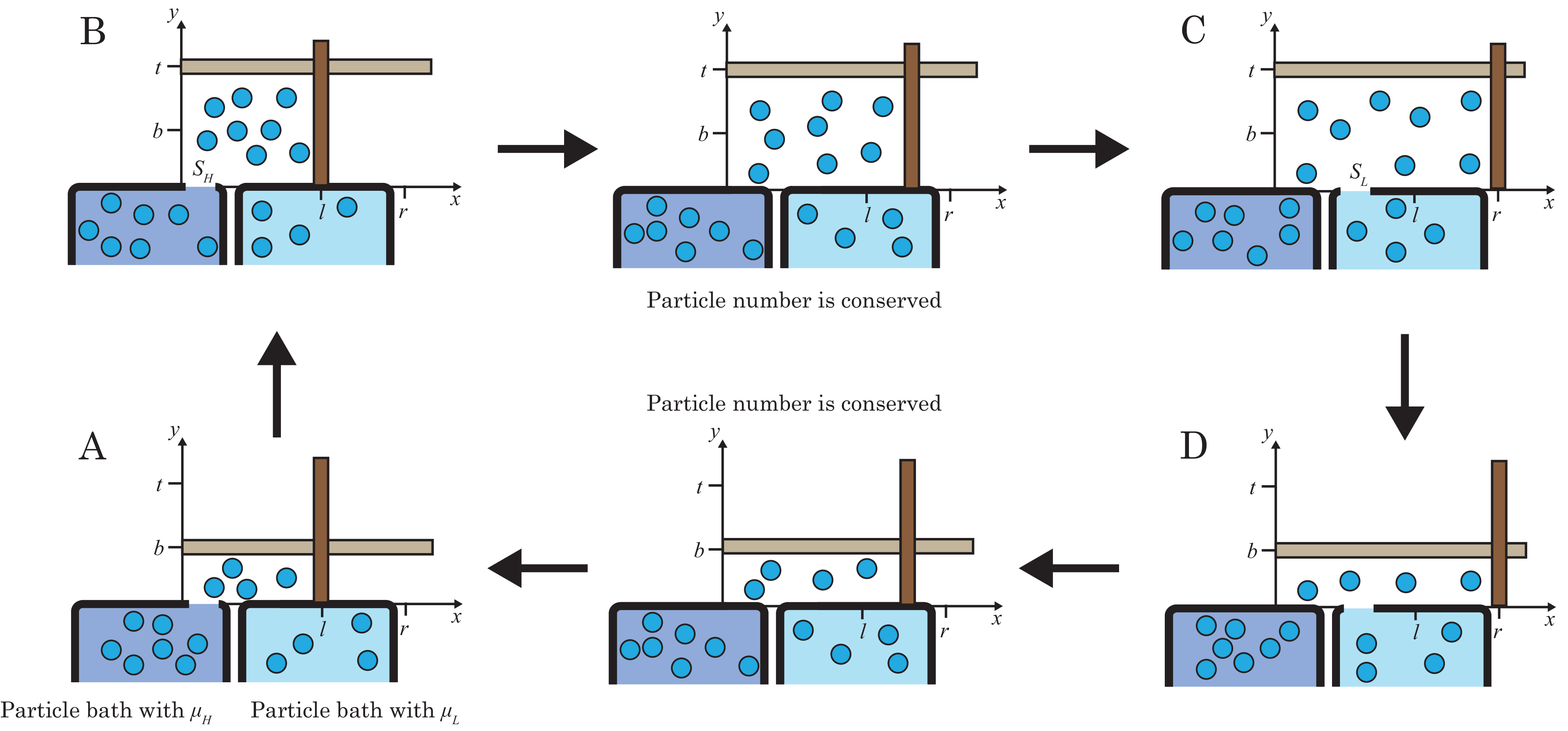}{
Schematic of the cyclic process of the autonomous Carnot engine.
$S_H$ ($S_L$) is a gateway between the engine and the bath with $\mu _H$ ($\mu _L$), which is open at $A$ and $B$ ($C$ and $D$) in this schematic.
Since $S_H$ is boundary, the dimension of $S_H$ is one dimension lower than that of the engine.
}{cycle}

We next discuss the autonomous attachment and detachment of particle baths.
Only when V-wall is at $r$ ($l$), the delta-function type energy barrier between the engine and the particle bath with $\mu _H$ ($\mu _L$) is removed and particles are exchanged between the engine and the bath.
This protocol is realized by using potentials as 
\balign{
U_H(x_i,X):=&\delta (x_i\in S_H)(1-\chi (x_l-\ep<X<x_l+\ep)), \\
U_L(x_i,X):=&\delta (x_i\in S_L)(1-\chi (x_r-\ep<X<x_r+\ep)),
}
where $S_H$ ($S_L$) is a border between the engine and the bath with $\mu _H$ ($\mu _L$) and $\delta (x_i\in S_H)$ takes infinity if $x_i$ is on $S_H$ and takes zero otherwise.
Note that such a protocol is realizable without any external control parameter.
By tuning the external force and potentials for walls, the engine moves as $A\to B\to C\to D\to A$ automatically and performs work against the force (see \fref{cycle}).

\

\begin{center}
{\bf \large Method to adding a new stochastic variable to satisfy conditions (a) and (b)}
\end{center}

\figin{12cm}{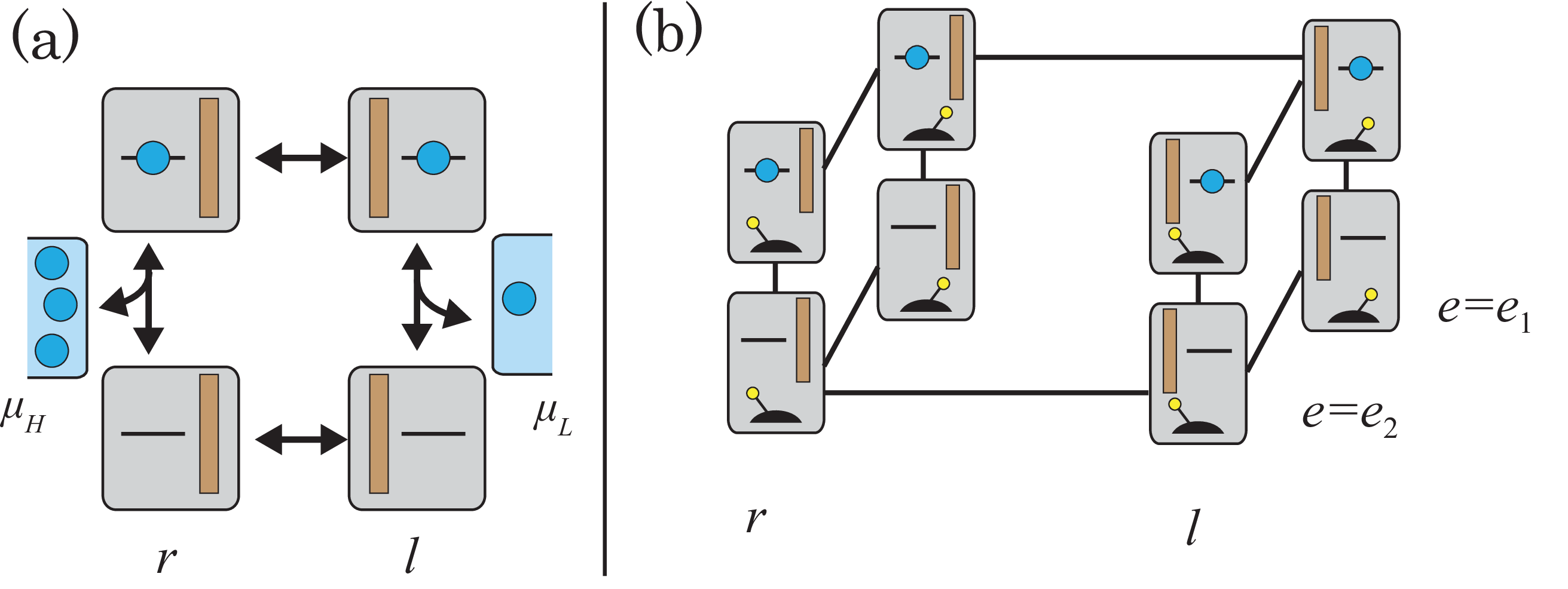}{
(a) State space of the whole system of autonomous Maxwell's demon.
When the engine is at $r$ ($l$), particles are exchanged with a particle bath with $\mu _H$ ($\mu _L$).
(b) Addition of a new stochastic variable $e\in \{ e_1,e_2\}$, which corresponds to a lever at right and left respectively.
}{info}

Autonomous engines discussed in Refs.~\cite{Feynman, Parrondo, Jarzynski, Sekimoto, Esposito, SS, SIKS, Mahan, Humphrey, Horvat, Sokolov, Seifert, Sanchez, Jordan, MJ, autoMJ, SMS} themselves do not satisfy the condition (a) or (b).
By taking an information engine~\cite{SS} as an example, we here demonstrate that how the addition of a new variable makes these engines satisfying the conditions (a) and (b).

The information engine is an isothermal engine with two particle baths with chemical potential $\mu _H$ and $\mu _L$.
The engine has two possible states $x\in \{ r,l\}$, and the engine possesses at most one particle (see \fref{info}.(a)).
The engine exchanges particles with a dense (dilute) particle bath when the state of the engine is $r$ ($l$).
The whole state of the system is determined by $(x,n)$.
We set the transition rates as satisfying
\balign{
\ln \frac{P((r,0)\to (r,1))}{P((r,1)\to (r,0))}=&\mu _H, \\
\ln \frac{P((l,0)\to (l,1))}{P((l,1)\to (l,0))}=&\mu _L, \\
\ln \frac{P((r,0)\to (l,0))}{P((l,0)\to (r,0))}=&E_0, \\
\ln \frac{P((r,1)\to (l,1))}{P((l,1)\to (r,1))}=&E_1
}
with $E_0\neq E_1$, which violates the condition (b).
Here, we normalized $\beta$ to 1.

We now introduce an additional stochastic variable $e\in \{ e_1,e_2\}$ to the engine (see \fref{info}.(b)).
The state of the {\it modified} engine is determined by $(x,e)$, and the state of the whole {\it modified} system is determined by $(x,e,n)$.
We then set the transition rates as
\eq{
P((x,n,e_1)\to (x,n,e_2))=P((x,n,e_2)\to (x,n,e_1))=\const
}
for any $x,n$, and
\balign{
P((x,0,e_1)\to (x,1,e_1))=P((x,0,e_2)\to (x,1,e_2))=&P((x,0)\to (x,1)), \\
P((x,1,e_1)\to (x,0,e_1))=P((x,1,e_2)\to (x,0,e_2))=&P((x,1)\to (x,0))
}
for $x=l,r$, and
\balign{
P((r,1,e_1)\to (l,1,e_1))=&2P((r,1)\to (l,1)), \\
P((l,1,e_1)\to (r,1,e_1))=&2P((l,1)\to (r,1)), \\
P((r,0,e_2)\to (l,0,e_2))=&2P((r,0)\to (l,0)), \\
P((l,0,e_2)\to (r,0,e_2))=&2P((l,0)\to (r,0)), \\
P((r,0,e_1)\to (l,0,e_1))=P((l,0,e_1)\to (r,0,e_1))=&0, \\
P((r,1,e_2)\to (l,1,e_2))=P((l,1,e_2)\to (r,1,e_2))=&0. 
}
It is easy to check that the modified engine satisfies the conditions (a) and (b).
Note that this addition procedure is different from the additional variable discussed in Ref.~\cite{SIKS}.
By setting the transition rates for $e$ much larger than that for $x$ and $n$, the coarse-graining of the fast variable $e$ leads to the original system written as $(x,n)$.

The addition procedures for other models are in a similar manner.
As an example, we briefly discuss the case of a discretized version of Feynman's ratchet~\cite{Jarzynski}.
The state of the engine $X_i$ is determined by the position of the pawl, which is discretized in two states, and the angle of the ratchet (see \fref{Feynman}).
The pawl is driven by the hot bath, and the ratchet is driven by the cold bath.
At any state, both the position of the pawl and the angle of the ratchet can change, which violates the condition (a).
We here add the new variable $e\in \{ e_H,e_L\}$, which divides the state of the engine $X_i$ into a hot bath state $X_{i,H}:=(X_i,e_H)$ and a cold bath state $X_{i,L}:=(X_i, e_L)$.
Then transitions between $X_{i,H}\lr X_{j,H}$ ($X_{i,L}\lr X_{j,L}$) are associated with only the hot (cold) bath, and thus the modified engine satisfies the conditions (a) and (b).

\figin{16cm}{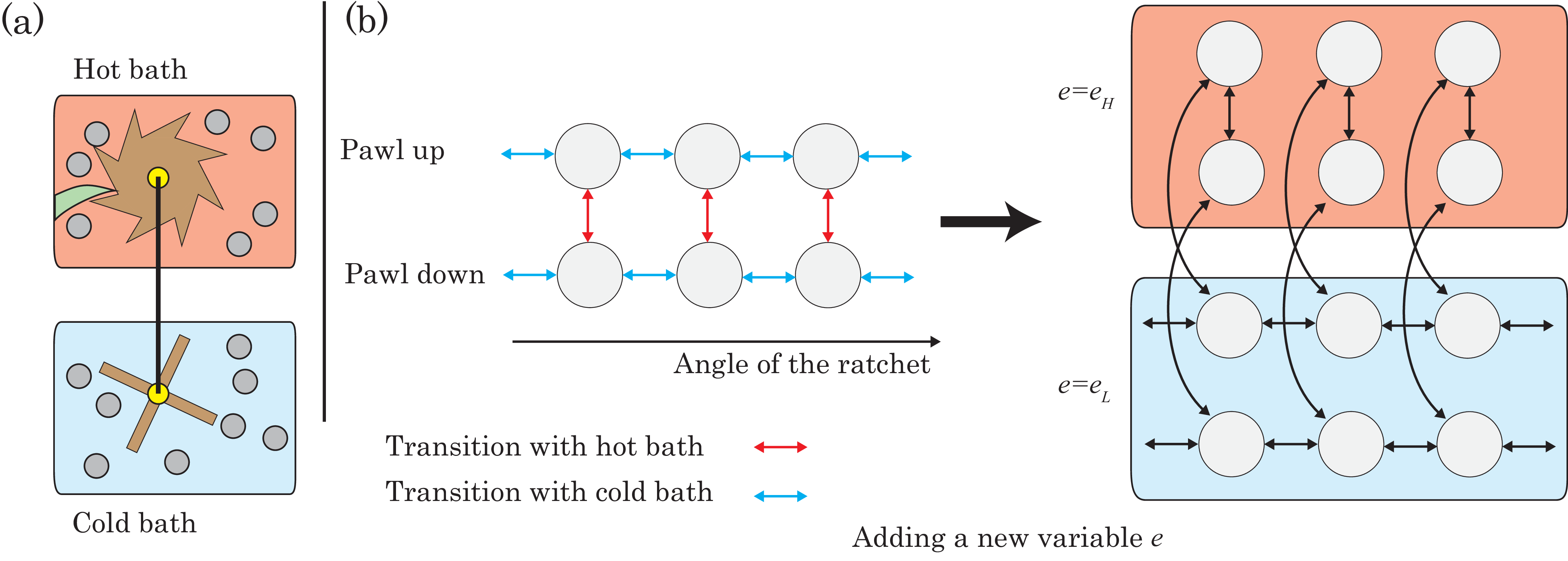}{
(a)  Schematic of the Feynman's ratchet and pawl~\cite{Feynman}.
The ratchet is driven by the cold bath and the pawl (i.e., green triangle) is driven by the hot bath.
(b) The state space of (discretized) Feynman's ratchet and pawl~\cite{Jarzynski} and that of modified one.
In the modified model, the additional parameter $e$ distinguishes whether the system is attached to the hot bath or the cold bath.
}{Feynman}

We finally make a remark on this procedure.
In the case of two heat baths we should introduce another heat bath which gives the stochastic nature of $e$.
Since the energy with $(X, e_H)$ and that with $(X, e_L)$ are the same, the third bath does not exchange any energy with hot or cold heat baths.

\

\begin{center}

{\bf \large Setting of CGACE which attains the CE}

\end{center}

We now write down the detailed setting of the CGACE which attains the CE.
The transition rates of the walls are given by 
\balign{
P(A\to D;n)&=\frac{k\cdot e^{-\beta (F(V_D,n)-E_{DA})}}{e^{-\beta F(V_A,n)}+e^{-\beta (F(V_D,n)-E_{DA})}}, \\
P(D\to A;n)&=\frac{k\cdot e^{-\beta F(V_A,n)}}{e^{-\beta F(V_A,n)}+e^{-\beta (F(V_D,n)-E_{DA})}}, \\
P(B\to C;n)&=\frac{k\cdot e^{-\beta (F(V_C,n)+E_{BC})}}{e^{-\beta F(V_B,n)}+e^{-\beta (F(V_C,n)+E_{BC})}}, \\
P(C\to B;n)&=\frac{k\cdot e^{-\beta F(V_B,n)}}{e^{-\beta F(V_B,n)}+e^{-\beta (F(V_C,n)+E_{BC})}}.
}
Such transition rates are realized by setting the potential of V-wall as \fref{pot}.
The position $A'$($D'$) is close to $A$($D$), where the distance of $A-A'$ ($D-D'$) are negligibly small.
We set $E_w$, energy barriers with $A-A'$ and $D-D'$, large enough for V-wall to equilibrate between $A'$ and $D'$ before leaving to $A$ or $D$.
V-wall climbs the barriers with the Arrhenius rule: $P(A\to A';n)=ke^{-\beta E_w}$.

We then set $V_A=V_0$, $V_B=r_1V_0$, $V_C=r_1r_2V_0$, $V_D=r_2V_0$, and take thermodynamic limit $V_0\to \infty$ with $r_1,r_2=O(1)$ and $E_{AB}, E_{BC}, E_{CD}, E_{DA}=O(V_0)$.
For sufficiently large $V_0$, the logarithms of the probability flux take the following asymptotic forms with certain domains of $n$:
\balign{
\ln j_{A\to D}(n)&\simeq {\beta (\mu _Hn-F(V_A,n))}+c_l &:n\gg \bar{n}_{AD}, \lb{jad} \\
\ln j_{B\to C}(n)&\simeq {\beta (\mu _Hn-F(V_B,n)-E_{AB})}+c_l &:n\gg \bar{n}_{BC}, \lb{jbc} \\
\ln j_{C\to B}(n)&\simeq {\beta (\mu _Ln-F(V_C,n)+E_{CD})}+c_r&: n\ll \bar{n}_{BC}, \lb{jcb} \\
\ln j_{D\to A}(n)&\simeq {\beta (\mu _Ln-F(V_D,n))}+c_r &: n\ll \bar{n}_{AD}. \lb{jda} 
}
Here, $c_l:=\ln ({kP_{\rm st}(l)}/{Z_{AB}})$ and $c_r:=\ln ({kP_{\rm st}(r)}/{Z_{CD}})$ are constants, and $\bar{n}_{AD}$ and $\bar{n}_{BC}$ are the solutions of 
\balign{
F(V_A,\bar{n}_{AD})-F(V_D,\bar{n}_{AD})+E_{DA}=&0, \lb{barAD} \\
F(V_B,\bar{n}_{BC})-F(V_C,\bar{n}_{BC})-E_{BC}=&0, \lb{barBC}
}
both of which are of $O(V_0)$.

\figin{5cm}{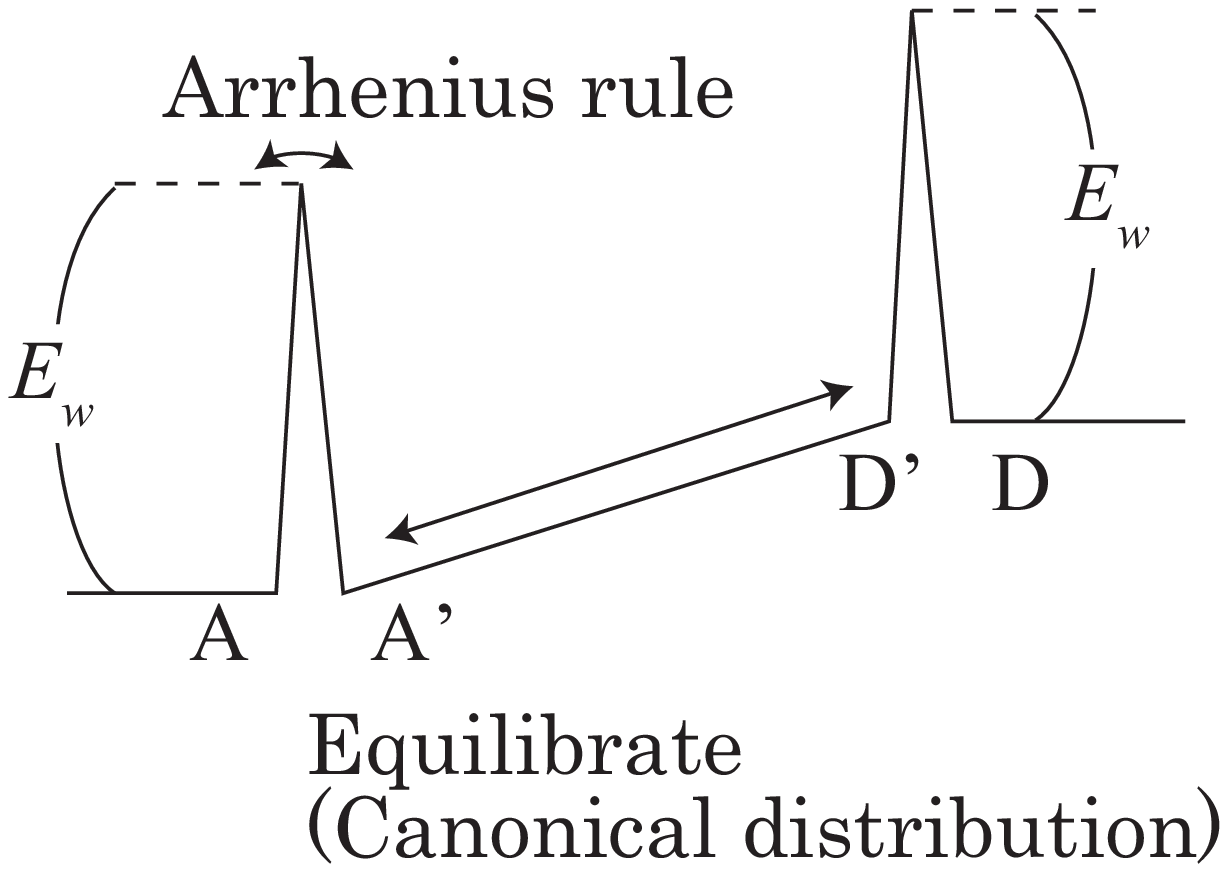}
{The potential landscape of V-wall.
Potential barriers with the height of $E_w$ exist at $A-A'$ and $D-D'$.
}{pot}

The relation \eqref{jad} is confirmed as follows.
The logarithm of $j_{A\to D}(n)$ is evaluated as
\balign{
\ln j_{A\to D}(n)= &c_l-\beta(F(V_A,n)-\mu _Hn)-\ln (1+e^{\beta (F(V_D,n)-F(V_A,n)-E_{DA})}) \nt \\
= &c_l-\beta(F(V_A,n)-\mu _Hn)-O(e^{\beta (F(V_D,n)-F(V_A,n)-E_{DA})}).
}
Here, for sufficiently large $V_0$, ${\beta (F(V_D,n)-F(V_A,n)-E_{DA})}$ with $n$ close to but larger than $\bar{n}_{AD}$ is evaluated as
\balign{
{\beta (F(V_D,n)-F(V_A,n)-E_{DA})}&=\beta (F(V_D,n)-F(V_A,n)-F(V_D, \bar{n}_{AD})+F(V_A,\bar{n}_{AD})) \nt \\
&\simeq \beta \( \ft{\frac{\del F}{\del n}}{r_2, \frac{\bar{n}_{AD}}{V_0}}-\ft{\frac{\del F}{\del n}}{1, \frac{\bar{n}_{AD}}{V_0}} \) (n-\bar{n}_{AD}) \nt \\
&=O(n-\bar{n}_{AD}),
}
whose sign is negative.
Therefore, by setting $n$ as $n-\bar{n}_{AD}=O({V_0}^{2/3})$, $e^{\beta (F(V_D,n)-F(V_A,n)-E_{DA})}\ll 1$ is satisfied, and since $\beta(F(V_A,n)-\mu _Hn)=O(V_0)$, the term $O(e^{\beta (F(V_D,n)-F(V_A,n)-E_{DA})})$ is negligible in our analyses.
Hence, we obtain \eref{jad}, and in a similar manner we also obtain Eqs.~\eqref{jbc}, \eqref{jcb}, and \eqref{jda}.
We note that the fluctuation of the realized particle number around $n_A^*$ when the transition $A\to D$ occurs is $O(\sqrt{n_A^*})=O(\sqrt{V_0})$, which is much smaller than $O({V_0}^{2/3})$.

Now we write down all parameters in terms of $V_0$ with fixed $r_1>1$ and $\delta >0$ (the procedure is summarized in \fref{flow}).
First, \eref{jad} tells $n_A^*$ as the solution of 
\eq{
\left. \frac{\del F}{\del n}\right|_{V_0, n_A^*}=\mu _H.
}
To achieve $n_D^*=n_A^*-{V_0}^{2/3}$ under \eref{jda}, $V_D$ is set as satisfying
\eq{
\left. \frac{\del F}{\del n}\right|_{V_D, n_A^*-{V_0}^{2/3}}=\mu _L,
}
where $V_D/V_A=:r_2=O(1)$ holds.
Owing to the extensiveness of the free energy, Eqs.~\eqref{jbc} and \eqref{jcb} yield
\eq{
\left. \frac{\del F}{\del n}\right|_{V_B, n_B^*}=\mu _H, \ \ \left. \frac{\del F}{\del n}\right|_{V_C, n_C^*}=\mu _L
}
with $n_B^*=r_1n_A^*$ and $n_C^*=r_1n_D^*$.
Next, we set $E_{DA}$ and $E_{BC}$ as satisfying Eqs.~\eqref{barAD} and \eqref{barBC} with $\bar{n}_{AD}=(n_A^*+n_D^*)/2$ and $\bar{n}_{BC}=(n_B^*+n_C^*)/2$:
\balign{
E_{DA}=&F\(V_D, \frac{n_A^*+n_D^*}{2}\)-F\( V_A, \frac{n_A^*+n_D^*}{2}\) ,  \lb{EDA} \\
E_{BC}=&F\( V_B, \frac{n_B^*+n_C^*}{2}\)-F\( V_C,\frac{n_B^*+n_C^*}{2}\) . \lb{EBC}
}
This setting of $E_{DA}$ and $E_{BC}$ ensures $n_A^*>\bar{n}_{AD}>n_D^*$,  $n_B^*>\bar{n}_{BC}>n_C^*$, and $|n_X^*-\bar{n}_{XX^-}|=O({V_0}^{2/3})$, $|n_{X^-}^*-\bar{n}_{XX^-}|=O({V_0}^{2/3})$, which confirm that $n_X^*$ is close to the discontinuous region of $G$.
Finally, we set $E_{AB}$ and $E_{CD}$ with a constant $\delta >0$ as follows:
\balign{
E_{AB}&=F(V_A,n_A^*)-F(V_B,n_B^*)+\mu _H(n_B^*-n_A^*)-\delta , \\
E_{CD}&=F(V_C,n_C^*)-F(V_D,n_D^*)+\mu _L(n_D^*-n_C^*)-\delta .
}

\figin{10cm}{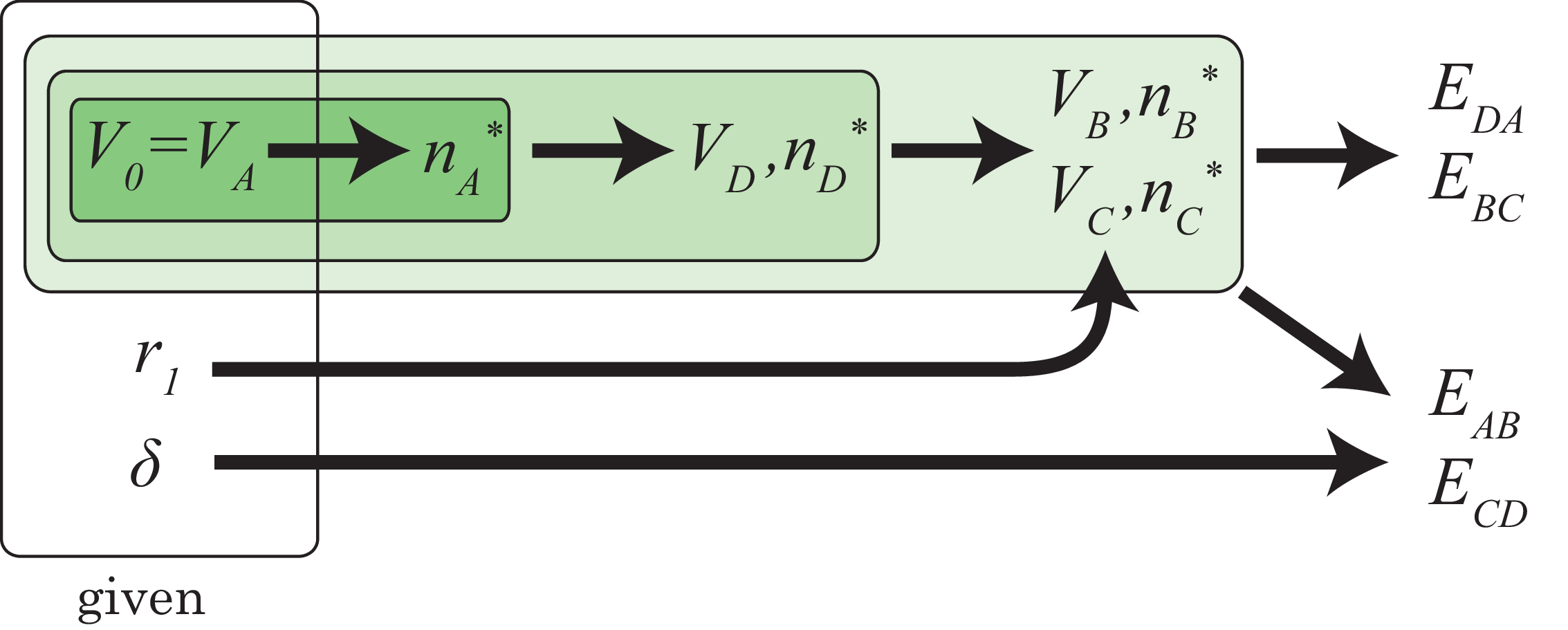}{
A flowchart of how the variables are determined by given $V_0$, $r_1$, and $\delta$.
For example, using $V_0=V_A$, $n_A^*$ is determined, and using $V_0$ and $n_A^*$, $V_D$ and $n_D^*$ are determined.
}{flow}

We now check that these settings satisfy Eqs.~\eqref{jcond1} and \eqref{jcond2}.
Owing to the stationary condition $j_{D\to A}(n_D^*)-j_{A\to D}(n_A^*)=j_{B\to C}(n_B^*)-j_{C\to B}(n_C^*)$, it is enough to check both $j_{A\to D}(n_A^*)<j_{B\to C}(n_B^*)$ and $j_{C\to B}(n_C^*)<j_{D\to A}(n_D^*)$.
Using Eqs.~\eqref{jad} and \eqref{jbc}, the logarithm of $j_{B\to C}(n_B^*)/j_{A\to D}(n_A^*)$ is calculated as
\balign{
\ln \frac{j_{B\to C}(n_B^*)}{j_{A\to D}(n_A^*)}
=\beta (\mu_H (n_B^*-n_A^*)-F(V_B,n_B^*)+F(V_A,n_A^*)-E_{AB})=\delta >0.
}
The logarithm of $j_{D\to A}(n_D^*)/j_{C\to B}(n_C^*)$ is calculated in a similar manner.
Hence, Eqs.~\eqref{jcond1} and \eqref{jcond2} are satisfied.

Finally, we evaluate $W_{\rm tot}$ and $C_\mu$.
First, $W_{\rm tot}$ is calculated as
\balign{
W_{\rm tot}
:=&F(V_A,n_A^*)-F(V_B,n_B^*)+F(V_C,n_C^*)-F(V_D,n_D^*)+\mu_H(n_B^*-n_A^*)+\mu_L(n_D^*-n_C^*)-2\delta \nt \\
&-F(V_A,\bar{N}_{AD})+F(V_D,\bar{n}_{AD})+F(V_B,\bar{n}_{BC})-F(V_C,\bar{n}_{BC}) .\lb{torqueAP}
}
Here, by using 
\eq{
\ft{\frac{\del F}{\del n}}{V_A, n_A^*}=\ft{\frac{\del F}{\del n}}{1, \frac{n_A^*}{V_0}}=\mu _H
}
and
\eq{
\frac{n_A^*}{V_0}-\frac{\bar{n}_{AD}}{V_0}=\frac{1}{2{V_0}^{1/3}},
}
we obtain
\eq{
F(V_A,n_A^*)-F(V_A,\bar{n}_{AD})=V_0\( F(1,\frac{n_A^*}{V_0})-F(1,\frac{\bar{n}_{AD}}{V_0})\) =V_0\cdot \( \mu _H\frac{1}{2{V_0}^{1/3}}+O\( \frac{1}{V_0}\) \) =O({V_0}^{2/3}).
}
In a similar manner, we obtain
\balign{
F(V_B,\bar{n}_{BC})-F(V_B,n_B^*)&=O({V_0}^{2/3}), \\
F(V_C,n_C^*)-F(V_C,\bar{n}_{BC})&=O({V_0}^{2/3}), \\
F(V_D,\bar{n}_{AD})-F(V_D,n_D^*)&=O({V_0}^{2/3}).
}
By combining $n_A^*-n_D^*={V_0}^{2/3}$ and $n_B^*-n_C^*={V_0}^{2/3}$ and above four relations, \eref{torqueAP} is evaluated as
\eqa{
W_{\rm tot} \geq (\mu _H-\mu _L)(n_B^*-n_D^*)+O({V_0}^{2/3}) .
}{Weva}
Next, $C_\mu$ is calculated as
\balign{
C_\mu 
=&\frac{(\mu _H -\mu _L)(e^\delta n_B^*+n_A^*-e^\delta n_D^*-n_C^*)}{e^\delta -1} +O(e^{-{V_0}^{2/3}}) \nt \\
=&(\mu _H-\mu _L)(n_B^*-n_D^*)+O({V_0}^{2/3}).
}
Since $n_B^*-n_D^*=O(V_0)$, the efficiency is evaluated as
\eq{
\eta :=\frac{W_{\rm tot}}{C_\mu }\geq 1-O\( \frac{1}{{V_0}^{1/3}}\) ,
}
which indicates attainability of the CE with $V_0\to \infty$.

\clearpage

\end{widetext}

\end{document}